\documentclass[twocolumn,showpacs,preprintnumbers,amsmath,amssymb]{revtex4}
\usepackage{graphicx}% Include figure files
\usepackage{dcolumn}% Align table columns on decimal point
\usepackage{bm}% bold math

\begin{document}

\title{The sudden birth and sudden death of thermal fidelity in a two-qubit system }
\author{Li-Jun Tian$^{1,2}$}
\author{Li-Guo Qin$^{1,2}$}\email{lgqin@shu.edu.cn}
\author{Ying Jiang$^{1,2}$}
\author{Hong-Biao Zhang$^{3}$}

\affiliation {$^1$Department of Physics, Shanghai University,
Shanghai, 200444, China\\$^2$Shanghai Key Lab for Astrophysics,
Shanghai, 200234, China\\$^3$Institute of Theoretical Physics,
Northeast Normal University, Changchun 130024, China}
\date{\today}

\begin{abstract}
We study the energy level crossings of the states and thermal
fidelity for a two-qubit system in the presence of a transverse and
inhomogeneous magnetic field. It is shown clearly the effects of the
anisotropic factor of the magnetic field through the contour figures
of energy level crossing in two subspaces, the isotropy subspace and
anisotropy subspace. We calculate the quantum fidelity between the
ground state and the state of the system at temperature $T$, and the
results show the strong effect of the anisotropic factor again. In
addition, by making use of the transition of Yangian generators in
the tensor product space, we study the evolution of the thermal
fidelity after the transition. The potential applications of Yangian
algebra, as a switch to turn on or off the fidelity, are proposed.

\end{abstract}


\maketitle

\section{Introduction}

Energy levels and eigenvalues of a Hamiltonian, especially the
ground state, play an important role in determining the properties
of a quantum system \cite{soz}. When the energy gap between the
ground and first excited levels tends to zero, the energy level
crossing will appear \cite{mva}. It is well known that the
degeneracy and the quantum phase transition lead to various peculiar
phenomena \cite{soz,mva,efj}. It is very important to investigate
level crossings to comprehend the quantum phase transition.
Bhattacharya and Raman presented an effective algebraic method for
finding level crossings \cite{mbc}. It is still important to explore
a way of studying level crossings and understand how eigenvalues of
the Hamiltonian change at crossing.

In order to detect phase transitions, the fidelity, as a new concept
originated from quantum information theory, has been put forward
\cite{pzn,htq,ssv}. Recently, the fidelity as a useful probe has
attracted much attention \cite{dsf,afa,hqz,pzl,ych,jrw}. In
addition, as a geometric measure, the fidelity may represent an
effective approach to identify quantum phase transitions
\cite{wly,lcv,htq2}. Moreover, it could be a good indicator to
energy level crossings \cite{pzp,mcp}. At present most efforts have
been devoted to the ground-state fidelity corresponding to slightly
different values of the controlling parameters. In this work, we
study the fidelity between the ground state and the state of the
system at temperature $T$ to explore the cases of the system in the
ground state.

In this paper, we study the two distinguishing qubits system in a
transverse and inhomogeneous magnetic field, and calculate the
energy level crossings and the quantum fidelity. For XY spin chain
model, the fidelity has been studied in the homogeneous magnetic
field \cite{pzn,hqz}. However, we will study the fidelity in the
non-homogeneous magnetic field, which acts on two different qubits.
By taking advantage of the nonuniform characteristic, we study its
effect to the energy level crossings and thermal fidelity, and find
new interesting physical phenomena.

In addition, due to the particular characteristic of Yangian algebra
to deal with physical models: symmetry and transition, Yangian
algebra method has been widely studied \cite{sle,tian,ljtq} in
recent years. People have found the Yangian symmetry in many
physical models \cite{Uglov,Kundu,bernard2}. It is also important
and helpful for exploring physical systems in terms of the
transition characteristic of Yangian. In this paper, by making use
of the transition effect of Yangian operators in the tensor product
space, we obtain the better regulation and control to the thermal
fidelity.

The paper is organized as follows: In Sec. II, we introduce a
two-qubit XY spin model Hamiltonian and divide the system into the
two subspaces: the isotropy and the anisotropy spaces. As an
important physical phenomenon, the energy level crossings are
studied for some models in detail in Sec. III. In Sec. IV, the
thermal fidelity between the ground state and the state of the
system at temperature $T$ is studied in the parametric space. The
effects of the anisotropic factor of the magnetic field to the
fidelity are explored. Sec. V presents the effects of Yangian
operators on the thermal fidelity. Finally, our conclusions are
given.

\section{The model Hamiltonian}

The Hamiltonian of a two-qubit XY spin model with an inhomogeneous
magnetic field in the $Z$ direction can be expressed as
\begin{eqnarray}
\label{lg1}
H=-\frac{1+\gamma}{2}\sigma^{x}_{1}\sigma^{x}_{2}-\frac{1-\gamma}{2}\sigma^{y}_{1}\sigma^{y}_{2}-B
(\sigma^{z}_{1}+\lambda \sigma^{z}_{2}),
\end{eqnarray}
where $\sigma^{\alpha}_{i}$ are the Pauli matrices of the i-th qubit
with $\alpha=x, y, z$ and $B$ is the strength of the external
magnetic field. $\lambda$ is an anisotropic factor of the magnetic
field acting on the second qubit and denotes the non-uniformity
degree of magnetic field. $\gamma$ is an anisotropic factor in the
interaction.

 The eigenstates and corresponding
eigenvalues of the Hamiltonian can be obtained as
\begin{eqnarray}
\label{lg2} &&H|\psi_{1}\rangle= E_{1}|\psi_{1}\rangle=-\sqrt{\xi^2+1}|\psi_{1}\rangle,\nonumber\\
&&H|\psi_{2}\rangle=E_{2}|\psi_{2}\rangle=\sqrt{\xi^2+1}|\psi_{2}\rangle,\nonumber\\
&& H|\psi_{3}\rangle=E_{3}|\psi_{3}\rangle=\sqrt{\eta^2+\gamma^2}|\psi_{3}\rangle,\nonumber\\
&&H|\psi_{4}\rangle=E_{4}|\psi_{4}\rangle=-\sqrt{\eta^2+\gamma^2}|\psi_{4}\rangle,
\end{eqnarray}
where
\begin{eqnarray}
&&|\psi_{1}\rangle = \frac{1}{N_1}(|10\rangle+a_1|01\rangle), \;\;|\psi_{2}\rangle = \frac{1}{N_2}( |10\rangle+a_2|01\rangle ),\nonumber\\
&&|\psi_{3}\rangle = \frac{1}{N_3}( \gamma |11\rangle +a_3
|00\rangle), |\psi_{4}\rangle = \frac{1}{N_4}(
\gamma |11\rangle + a_4|00\rangle).\nonumber\\
\end{eqnarray}
Here $a_1=\xi+\sqrt{\xi^2+1}$, $a_2=\xi-\sqrt{\xi^2+1}$,
$a_3=\eta-\sqrt{\eta^2+\gamma^2}$ and
$a_4=\eta+\sqrt{\eta^2+\gamma^2}$, where $\xi=B(1-\lambda)$,
$\eta=B(1+\lambda)$, while $|0\rangle$ stands for spin down and
$|1\rangle$ stands for spin up. $N_{i}$ is the normalization
coefficient of $|\psi_{i}\rangle$ ($i=1,2,3,4$).

The Hamiltonian $H$ is acquired a more effective form by rewriting
Eq. ({\ref{lg1}}) in the matrix form \cite{gsp,soh},
\begin{eqnarray}
\label{llg1} H = H_{e} \oplus H_{o},
\end{eqnarray}
where $H_{e}=-\left(\begin {array}{cc}\eta & \gamma \\ \gamma &
-\eta\end {array}\right)$ and $H_{o}=-\left(\begin {array}{cc}\xi &
1\\ 1 & -\xi\end {array}\right)$, acting in an isotropy subspace
spanned by
$\{|\uparrow\uparrow\rangle,|\downarrow\downarrow\rangle\}$ and in
an anisotropy subspace by
$\{|\uparrow\downarrow\rangle,|\downarrow\uparrow\rangle\}$,
respectively. From Eqs. ({\ref{lg2}}) and ({\ref{llg1}}), we can
know $E_{3,4}$ and $|\psi_{3,4}\rangle$ are the eigenvalues and
eigenvectors of Hamiltonian $H_{e}$ respectively, and
$|\psi_{4}\rangle$ is the ground states in the isotropy subspace.
Similarly, the eigenvalues $E_{1,2}$ and the eigenvectors
$|\psi_{1,2}\rangle$ correspond to the eigenvalues and eigenvectors
of the Hamiltonian $H_{o}$ with the ground state $|\psi_{1}\rangle$
in the anisotropy subspace. Based on the two subspaces, which are
also closely related to Yangian algebra, the energy level crossings
and the thermal fidelity will be studied in detail as follows.

\section{the energy level crossings of the ground state}
From Eq. ({\ref{lg2}}), we learn that the ground state can only be
$|\psi_{1}\rangle$ or $|\psi_{4}\rangle$. However, by changing the
parameters, the rules of the ground state may be exchanged. When the
energy gap $_\Delta E=E_{1}-E_{4}$ is less than zero, it indicates
that $|\psi_{1}\rangle$ is the ground state; while $_\Delta
E=E_{1}-E_{4}>0$, it shows that $|\psi_{4}\rangle$ is the ground
state. However, the ground states are the degenerate states of
$|\psi_{1}\rangle$ and $|\psi_{4}\rangle$ for $_\Delta
E=E_{1}-E_{4}=0$. In order to express clearly the evolution of the
energy gap $_\Delta E$ in the parameter space, we are going to
investigate them graphically.

\begin{figure}[h]
\includegraphics[angle=0,width=9cm]{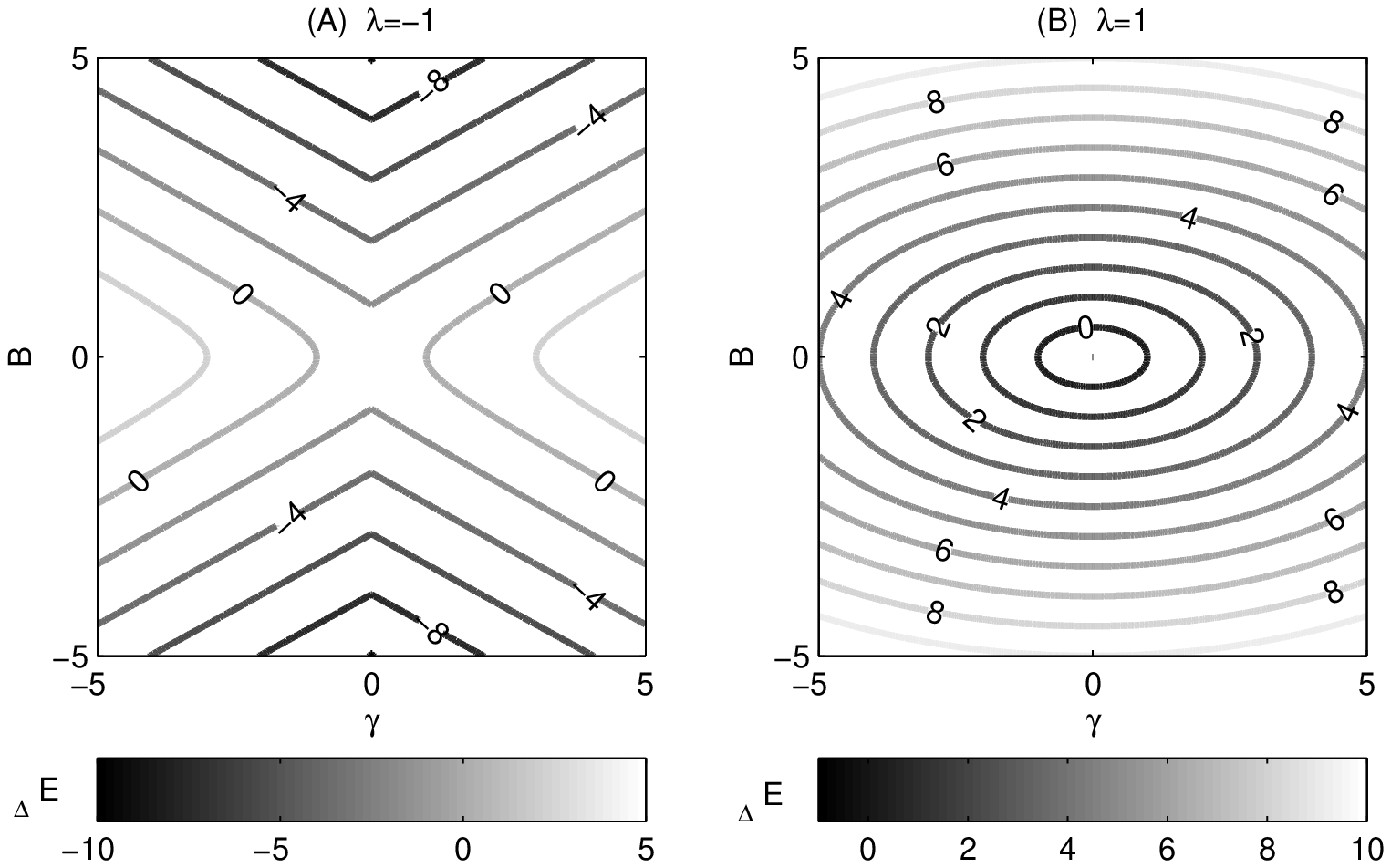}
\includegraphics[angle=0,width=9cm]{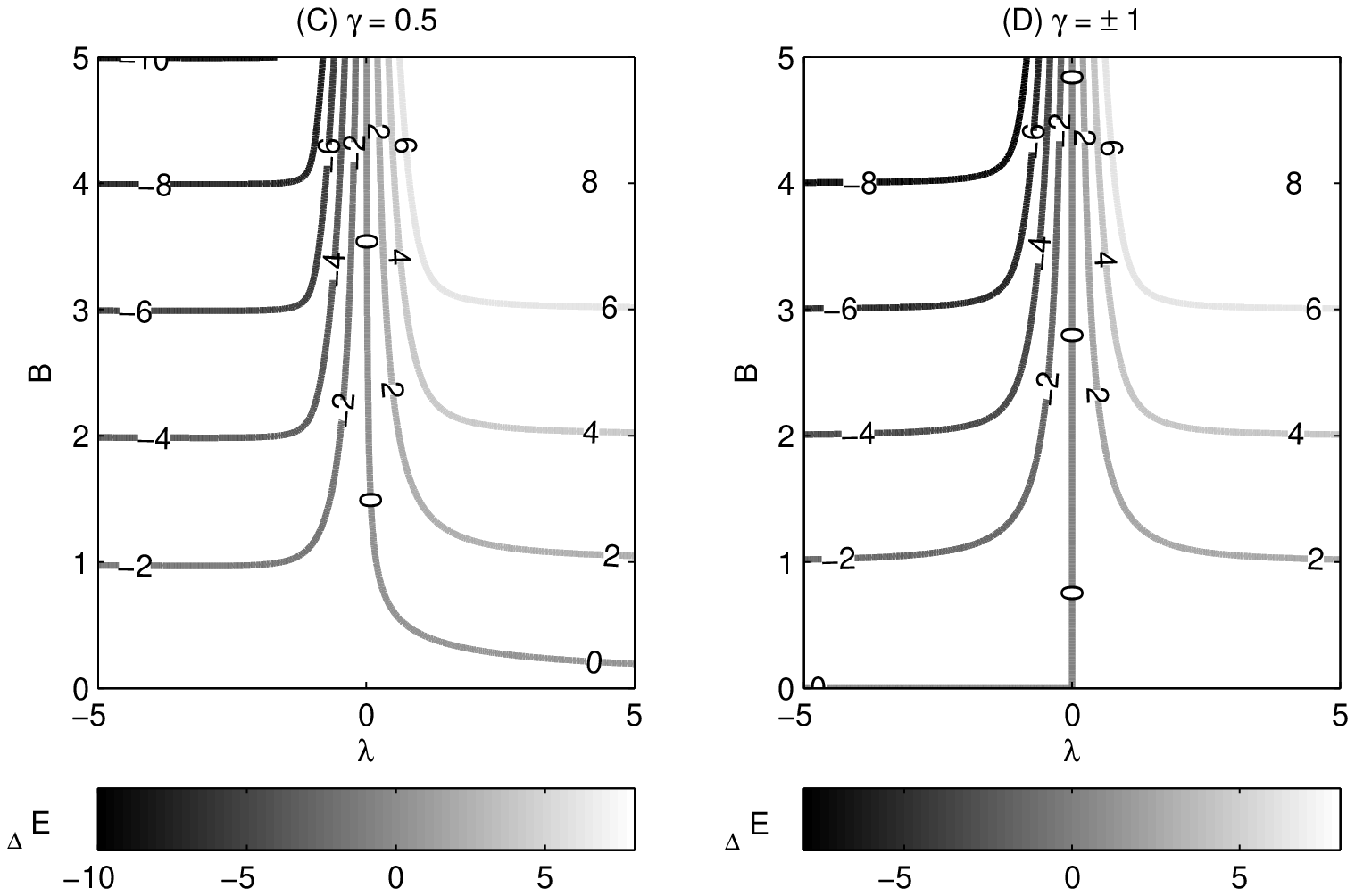}
\caption{(Color online) The contour plots of the energy gap. The
model is in the inhomogeneous magnetic field for $\lambda=-1$ in
(A), and the homogeneous magnetic field for $\lambda=1$ in (B). The
model is the ordinary $XY$ model in (C), and reduces the Ising model
in (D). The digital labels are the value of $|_\Delta E|$ in contour
graphs. }
\end{figure}

Figure 1 shows the contour plots of the energy gap as a function of
the magnetic field strength $B$ and the anisotropic factor $\lambda$
or $\gamma$ in the different cases. The energy level crossing of the
system occurs at the zero-contour lines, on which $E_{1}=E_{4}$,
namely, the energy level of the ground state happens to mutate at
these points. It is clear that the ground state changes abruptly at
$|_\Delta E|=0$ in Fig. 1.

For the two different magnetic fields: non-homogeneous for
$\lambda=-1$ or homogeneous for $\lambda=1$, the contours of
$_\Delta E$ versus $B$ and $\gamma$ have obvious differentiations.
In the Fig. 1(A), the contours of $_\Delta E<0$ are to be in the
upper and under polylines, which indicate that the ground state is
$|\psi_{1}\rangle$. while those of the left and right sides are
larger or equal to $0$, which show the ground state is
$|\psi_{4}\rangle$ or the degenerate state. In Fig. 1(B), the
contours are a series of concentric ellipses and the energy gap
becomes bigger as the radius increases. Especially, $_\Delta E<0$ is
within the zero contour, which indicate $|\psi_{1}\rangle$ is the
ground state. Without the zero contour, $_\Delta E>0$ indicates that
the ground state is $|\psi_{4}\rangle$. By comparing, we can find
that the probability of $|\psi_{1}\rangle$ as the ground state, the
areas of $_\Delta E<0$ in the ($\gamma$, $B$) plan, is comparatively
large for the inhomogeneous magnetic field, and almost disappears
completely in the case of the homogeneous magnetic field.

Without loss of generality, for $\gamma=0.5$, the system is the XY
model, and only has the coupling strength in the X or Y direction
for $\gamma=\pm 1$, that to say, the XY model becomes the
transverse-field Ising mode. Figs. 1(C) and (D) show the contour
graphs of $_\Delta E$ in the models. In Figs. 1(C) and (D), the
positive and negative of $_\Delta E$ have the same indications with
Figs. 1(A) and (B). In the case of $\gamma=0.5$, the probability of
$|\psi_{1}\rangle$ as the ground state is bigger than
$|\psi_{4}\rangle$. However, for $\gamma=\pm1$, the zero contour is
a straight line at $\lambda=0$ and the contour plots are bilateral.
The probability of $|\psi_{1}\rangle$ as the ground state is equal
to the case of $|\psi_{4}\rangle$. It is worth noting that
$|\psi_{4}\rangle$ is in the isotropy subspace and
$|\psi_{1}\rangle$ is in the anisotropy subspace. If the ground
state of the system, $|\psi_{1}\rangle$ or $|\psi_{4}\rangle$ is
identified, one can confirm the system is in the isotropy or
anisotropy subspace at zero temperature. From Fig. 1, one can obtain
the different ground state by adjusting the parameters.

\section{evolution of the fidelity}
Let us now turn our attention to probe the probability of the system
in the ground state of the anisotropy subspace. As a usable method,
the fidelity is defined by \cite{rjj,yzg}
\begin{eqnarray}
\label{lg3} F= \langle\psi|\rho|\psi\rangle,
\end{eqnarray}
where $|\psi\rangle$ is the pure state and $\rho$ is the density
matrix of the system or state. In our problem, we will study the
fidelity between the ground state of the anisotropy subspace and the
state of the system at temperature $T$. The density matrix $\rho(T)$
of the model in the thermal-equilibrium state at temperature $T$ is
given by
\begin{eqnarray}
\label{lg4} \rho= \sum^{4}_{i=1} p_{i}|\psi_i\rangle\langle\psi_i|,
\end{eqnarray}
where $p_i= \exp(-E_i/kT)/Z$ are the probability distributions and
the partition function $Z=Tr[exp(-H/kT)]$. For simplicity, we set
$k=1$ in all following equations. Based on $\{|11\rangle,
|10\rangle, |01\rangle, |00\rangle\}$, $\rho$ of the system is a
standard $X$ state given by
\begin{eqnarray}
 \label{lg5}\rho(T)=\frac{1}{Z} \left( \begin {array}{cccc}
v_1&0&0&u\\\noalign{\medskip}0&w_1&y&0\\\noalign{\medskip}0
&y&w_2&0\\\noalign{\medskip}u&0&0&v_2
\end {array} \right).
\end{eqnarray}
Here $v_1=(b_3+b_4)\gamma^2$, $v_2=b_3a_3^2+b_4a_4^2$,
$u=(b_3a_3+b_4a_4)\gamma$, $y=b_1a_1+b_2a_2$, $w_1=b_1+b_2$ and
$w_2=b_1a_1^2+b_2a_2^2$, where $b_i=\exp(-E_i/T)/N_i$. By making use
of Eq. (\ref{lg3}), the fidelity between the state of the system at
temperature $T$ and the ground state $|\psi_1\rangle$ of the
anisotropy subspace can be expressed as
\begin{eqnarray}
\label{lg6} F=\frac{w_1+2ya_1+w_2a_1^2}{Z(1+a_1^2)}.
\end{eqnarray}

\begin{figure}
\includegraphics[angle=0,width=8cm]{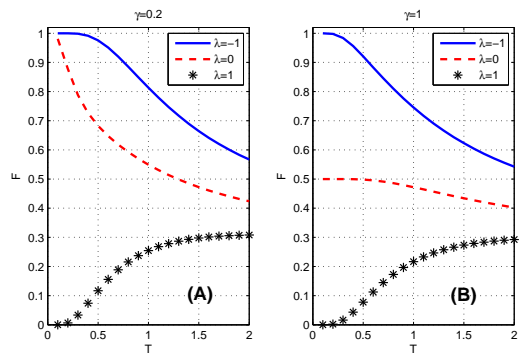}
\caption{(Color online) (A) and (B) show respectively $F$ as a
function of the temperature $T$ with the three different magnetic
fields in the two cases of $\gamma =0.2$ and $\gamma =1$.}
\end{figure}

\begin{figure}
\includegraphics[angle=0,width=8cm]{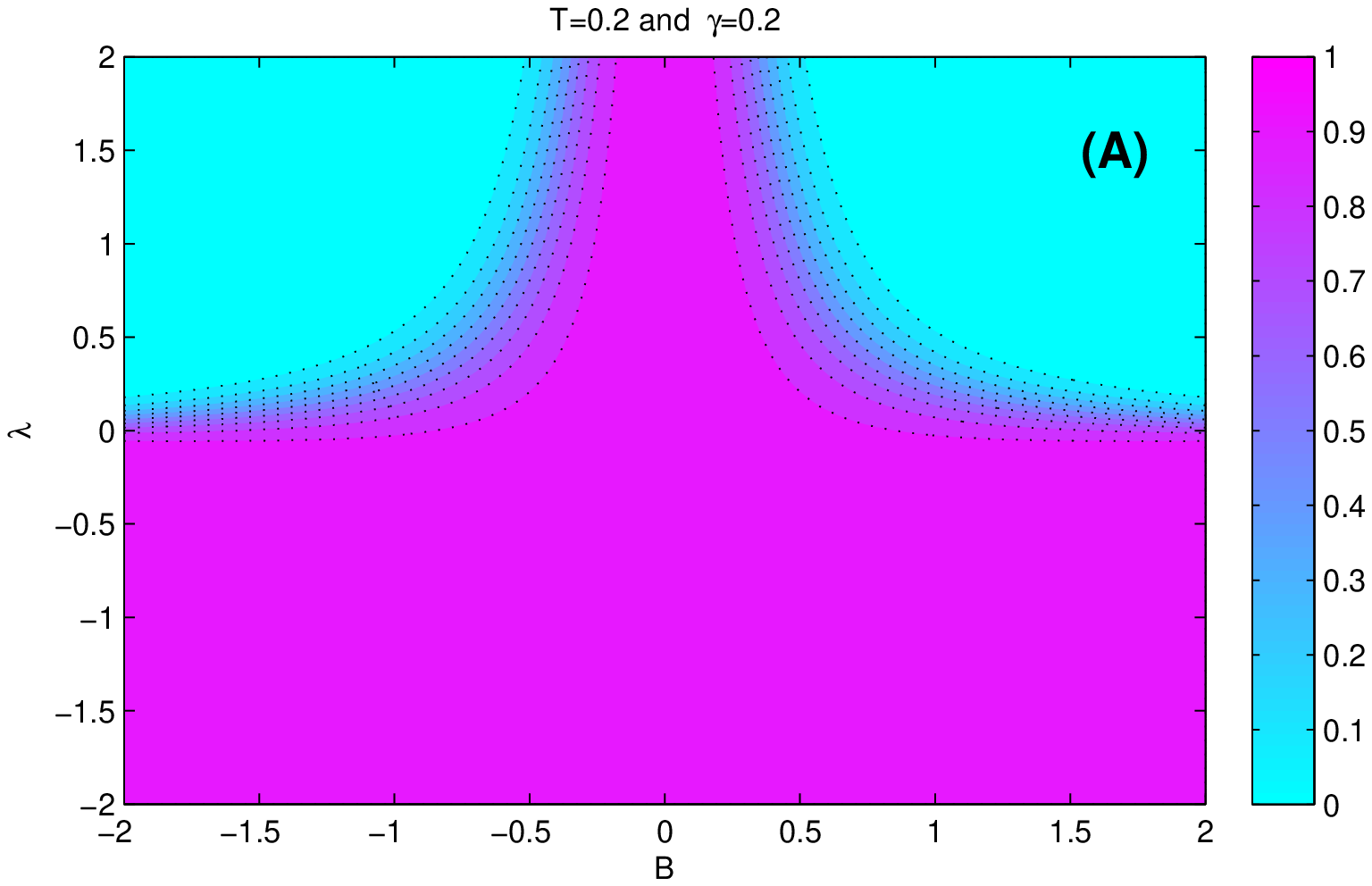}
\includegraphics[angle=0,width=8cm]{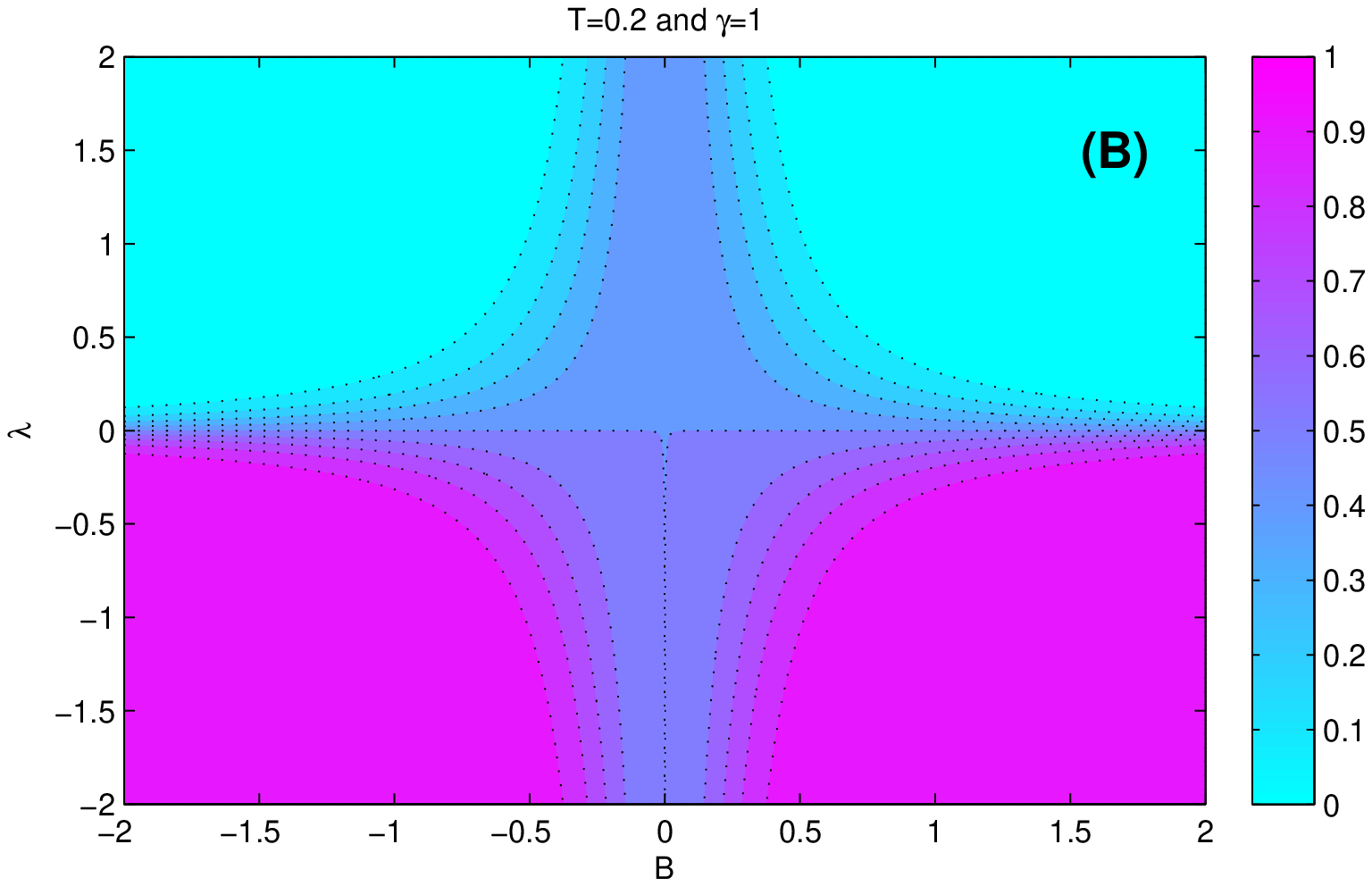}
\caption{(Color online) The contours of the thermal fidelity in
($B$, $\lambda$), space for $\gamma=0.2$ and $\gamma=1$ at $T=0.2$.}
\end{figure}

In Fig. 2, we show the evolution behavior of the thermal fidelity vs
$T$ with three different magnetic fields in the XY model
($\gamma=0.2$) and Ising model ($\gamma=1$). Under the conditions of
Fig. 2, the ground state is $|\psi_1\rangle$ for the inhomogeneous
magnetic field and $|\psi_4\rangle$ for the homogeneous magnetic
field. At zero temperature, the system will be in the ground state.
If the system is in $|\psi_1\rangle$, the overlap between the state
of the system and $|\psi_1\rangle$ is maximum, so one can obtain
$F=1$ for $\lambda=-1$. On the contrary, $F=0$ for $\lambda=1$
because $|\psi_4\rangle$ and $|\psi_1\rangle$ are orthogonal. In the
cases of $\lambda=0$, one get $F=1$ in Fig. 2(A), because
$|\psi_1\rangle$ is still the ground state for $\gamma=0.2$;
however, $F=0.5$ in Fig. 2(B), because the ground states become the
degeneracy states of $|\psi_1\rangle$ and $|\psi_4\rangle$ for
$\gamma=1$. As $T$ increases, the ground state mixes with the
excited states, the change trends of the fidelity are to be shown in
the illustration.

Figure. 3 displays the fidelity contours in ( $B$$,\lambda$) space
for $\gamma=0.2$ and $\gamma=1$ at $T=0.2$ respectively. The
fidelities are very susceptible to $\lambda$ and vary dramatically
jumping from zero to maximal around the critical points. The high
thermal fidelities mainly appear in the third and the fourth
quadrants of the ($B$, $\lambda$) space. Especially for the reversed
inhomogeneous magnetic field $\lambda <0$ and near $B=0$, the
fidelity is enhanced as shown in Fig. 3(A). When the XY model
reduces to the Ising model, by comparing Fig. 3(A) with Fig. 3(B),
one can find the fidelity declines around $B=0$ because of the
degeneracy as displayed in Fig. 3(B).

\begin{figure}
\includegraphics[angle=0,width=8cm]{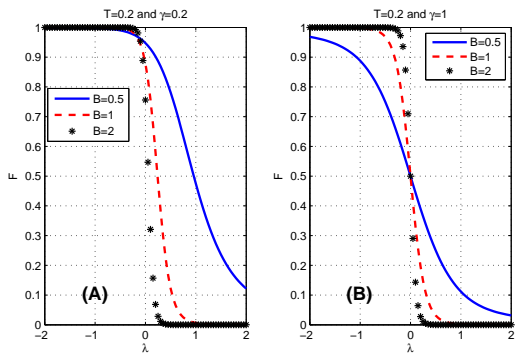}
\caption{(Color online) Two-dimensional plots of the fidelity as a
function of $\lambda$ for the three different magnetic fields in the
cases of $\gamma=0.2$ in (A) and $\gamma=1$ in (B) at $T=0.2$.}
\end{figure}

In order to illustrate the evolution of the thermal fidelity around
the critical points, we plot the graphs of $F$ vs $\lambda$ for
$\gamma=0.2$ in Fig. 4(A) and for $\gamma=1$ in Fig. 4(B) at
$T=0.2$. At low temperature, the system is primarily covered the
ground state. According to the expressions of energy levels in Eq.
({\ref{lg2}}), the ground state is $|\psi_1\rangle$ for $|\gamma|
\leq1$ and $\lambda <0$, and $|\psi_4\rangle$ is the ground state in
the other cases. We have $F=1$ for $|\psi_1\rangle$ as the ground
state, $F=0$ for $|\psi_4\rangle$ as the ground state, and $0<F<1$
for the mixed states of $|\psi_1\rangle$ and $|\psi_4\rangle$ as the
ground state in Fig. 4. In Fig. 4(A), as $\lambda$ increases, $F$
quickly decays from $1$ to $0$ around the critical point
($\lambda=0$). It is worth noting that the probabilities of
$|\psi_1\rangle$ and $|\psi_4\rangle$ in the ground state are
respectively $50\%$ for $\gamma = \pm1$ and $\lambda=0$ with the
random $B$, so the three curves cross at a point of $F=0.5$ and
$\lambda=0$, as shown in Fig. 4(B). In addition, especially $F=0$,
the fidelity is rapid death because $|\psi_1\rangle$ and
$|\psi_4\rangle$ are orthogonal. Due to the sharp change of the
fidelity as a function of $\lambda$, $\lambda$ can be used as a
switch to turn on or off the fidelity. For these properties,
possible applications are expected to quantum logical gate.

\section{evolution of the fidelity with the transition effect of Yangian generators}

Since Yangian algebra was presented by Drinfeld in 1985, it has
attracted much attention. Not only can Yangian algebra describe the
symmetry of quantum integrable models, but also can present the
transitions of the states between different weights beyond the Lie
algebra. For a bi-spin system, the realization of Yangian $Y(sl(2))$
is taken the form of \cite{2Ge1,ami}
\begin{eqnarray}
\label{definition
ij}&&{\bf{\emph{\textbf{I}}}}={\bf{\emph{\textbf{S}}}}={\bf{\emph{\textbf{S}}}_1}+{\bf{\textbf{S}}_2},\nonumber\\
&&{\bf{\emph{\textbf{J}}}}=\mu{\bf{\emph{\textbf{S}}}}_1}+\nu{\bf{\emph{\textbf{S}}_2}+i\lambda{\bf{\emph{\textbf{S}}_1}\times{\bf{\emph{\textbf{S}}}}_2},
\end{eqnarray}
where ${\bf{\emph{\textbf{S}}_1}}$, ${\bf{\emph{\textbf{S}}_2}}$ are
the spin-$\frac{1}{2}$ operators and $\mu$, $\nu$ and $\lambda$ are
arbitrary parameters. ${I}_\pm={I_1}{\pm}i{I_2}$ and
${J}_\pm={J_1}{\pm}i{J_2}$. ${\bf{\emph{\textbf{I}}}}$ is the total
spin operator satisfying $[I_i^a,I_j^b]=i \epsilon_{abc} I_i^c
{\delta}_{i j}, (i, j=1, 2)$, namely, the generators of the Lie
algebra. By making use of the transition, $\bf{\emph{\textbf{I}}}$
can only transit between the states with the same weight. However,
$\bf{\emph{\textbf{J}}}$ can transit mutually to different sites or
particles and jump between states with different weights
\cite{ljtq}.
\begin{figure}
\includegraphics[angle=0,width=8cm]{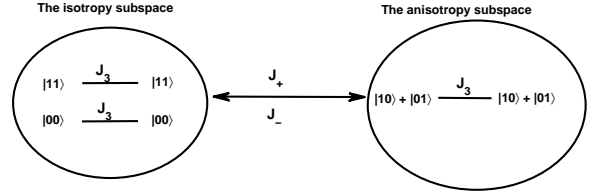}
\caption{(Color online) The transition graph of Yangian generators
between states in two subspaces $H_{e}$ and $H_{o}$.}
\end{figure}

Fig. 5 sketches the transition characteristic of the Yangian algebra
in the two subspaces. $J_3$ only makes the transitions of states
inside the subspace $H_e$ and $H_o$ respectively, however $J_{\pm}$
can transit from a state of one subspace to a state of another
subspace. This important link between the two subspaces and Yangian
algebra is suitable for any bi-spin systems, not only the XY model.
The effect of Yangian operators will contribute to the boost of the
high fidelity.

To illustrate the effect of Yangian algebra on the fidelity, we
discuss it as follows. By the transition characteristic of the
Yangian algebra, first let us act the transition operators $J_{\pm}$
on the initial state $|\psi_1\rangle$, the corresponding final
states are given by

\begin{eqnarray}
\label{lg7} |\psi^{'}_1\rangle = J_+ |\psi_1\rangle = |11\rangle,
\;\;\;\; |\psi^{''}_1\rangle = J_- |\psi_1\rangle = |00\rangle
\end{eqnarray}
with the normalization condition. How is the overlap between the
final states and the state of the system at temperature $T$? By
making use of Eq. (\ref{lg3}), the fidelity between the two final
states and the state of the system at temperature $T$ can be
obtained respectively as
\begin{eqnarray}
\label{lg8} F^{'}=\frac{w_1}{Z}, \;\;\;\;F^{''}=\frac{w_2}{Z}.
\end{eqnarray}

\begin{figure}
\includegraphics[angle=0,width=8cm]{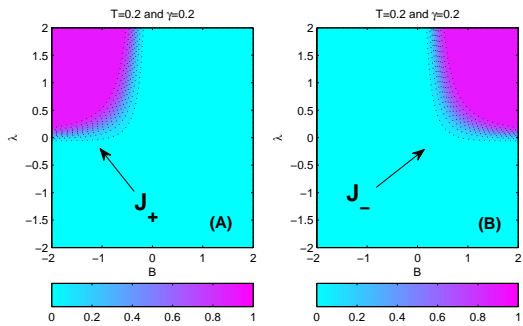}
\caption{(Color online) Two-dimensional contours of the fidelity
under the transition effect of $J_{\pm}$ for $\gamma =0.2$ in
($\lambda$, $B$) space.}
\end{figure}
\begin{figure}
\includegraphics[angle=0,width=8cm]{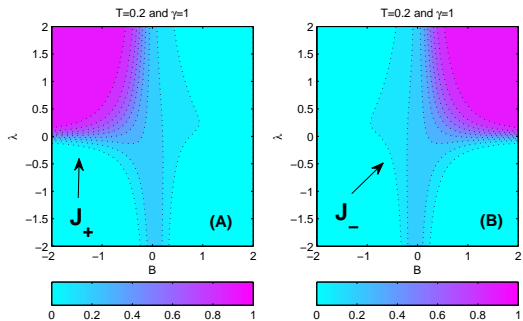}
\caption{(Color online) Two-dimensional contours of the fidelity
under the transition effect of $J_{\pm}$ for $\gamma =1$ in
($\lambda$, $B$) space.}
\end{figure}

We plot the contours of the fidelity after the transition for
$\gamma=0.2$ and $\gamma=1$ in Figs. 6 and 7. By comparing Figs. 6
and 7 with Fig. 3, one can find that by making use of the operator
$J_{+}$, the high fidelity transits from $\lambda <0$ to $\lambda
>0$ and $B<0$, and vanishes when $B>0$; accordingly, it only was centered on $\lambda>0$
and $B>0$ by taking advantage of $J_{-}$ in ($\lambda$, $B$) space.
By comparing the effect before and after the transition of Yangian
algebra from Figs. 3(A) and 6, we can clearly find that the high
fidelities are complemented in addition to the partial regular
extinction. It is the similar case by comparing Figs. 3(B) and 7.
High fidelity transits from the third and the fourth quadrants to
the second quadrant by the operator $J_{+}$ and to the first
quadrant by $J_{-}$ in ($B$, $\lambda$) space. For $\gamma=0.2$, the
ground state is $|\psi_1\rangle$, and there is not degeneracy in the
case of $B=0$. The initial state is orthogonal to the final states
$|\psi^{'}_1\rangle$ and $|\psi^{''}_1\rangle$, namely, the overlap
between two final states and the ground state passes out. Hence
$F^{'}=F^{''}=0$, and there is not a long tail as shown in Fig. 6.
However, the degeneracy of the ground states will appear for
$\gamma=1$, namely, the ground state is made up of $|\psi_1\rangle$
and $|\psi_4\rangle$. The overlap between two final states and the
ground state is not zero. So one can see a long tail around $B=0$ in
Fig. 7. Whatever, the high fidelities are concentrated more on the
one quadrant of a plan and are almost sudden death in the others
through the transition effect of Yangian generators in the tensor
product space. These results show that Yangian has the better hand,
as a switch, to turn on or off the fidelity. To the better
properties, possible applications are expected.
\section {Conclusions}
In summary, we have investigated the properties of the energy level
crossings of the ground states in the two subspaces and the
evolution of the thermal fidelity in the two-qubit system with a
transverse non-homogeneous magnetic field respectively. We also
discussed the effects of the quantum algebra on the states and high
fidelity in the isotropy subspace and anisotropy subspace. For
important parameters, we plotted the contour figures of energy gap,
with the help of which the ground state can be distinguished and the
level crossings are easily recognized. The thermal fidelity is
completely calculated in both the XY model ($\gamma =0.2$) and Ising
model ($\gamma=1$). In the two models the evolution tendencies of
thermal fidelity versus the temperature are uniform when
$\lambda=\pm 1$; but when $\lambda=0$ at $T=0$, the value of the $F$
in XY model is two times than one in Ising model, because the energy
level crossing appears. At $T=0.2$ the contours of fidelity versus
the parameters $\lambda$ and $B$ show $F$ rapid death around
$\lambda=0$ besides minimal region like $\delta$ function. Therefore
the anisotropic factor of the magnetic field $\lambda$, one of the
focus in this paper, can be modulated to control the thermal
fidelity. In addition, the thermal fidelity figures between the two
final states and the state of the system at temperature $T$ are
obtained by the transition effect of Yangian generators, and show
that the high fidelity only appears in the one branch ($\lambda>0$)
of the equilateral hyperbola in the ($B$, $\lambda$) plane.
Moreover, Yangian algebra plays important roles in enhancing the
high $F$ and making it more concentrate on the one quarter plane. So
to speak, we present a more optimal proposal to turn on or off the
fidelity as a switch.

These results show that Yangian algebra shed new light on fidelity
controlling and the effect of the Yangian algebra is tremendous to
deal with some physical models.

\begin{acknowledgments}

This work is partly supported by the NSF of China (Grant No.
11075101), Shanghai Leading Academic Discipline Project (Project No.
S30105), and Shanghai Research Foundation (Grant No. 07d222020).

\end{acknowledgments}

\end{document}